\documentclass[reprint,superscriptaddress,amappendixath,amssymb,aps]{revtex4-2}
\usepackage{graphicx}
\usepackage{dcolumn}
\usepackage{bm}
\usepackage{amsmath}
\usepackage{qcircuit}
\usepackage{algpseudocode}
\usepackage{algorithm}
\usepackage{subfigure}
\usepackage[pass]{geometry}
\usepackage{float}
\makeatletter
\renewcommand*\env@matrix[1][\arraystretch]{%
\edef\arraystretch{#1}%
\hskip -\arraycolsep
\let\@ifnextchar\new@ifnextchar
\array{*\c@MaxMatrixCols c}}
\makeatother
\begin{document}
\preprint{APS/123-QED}
\title{Hermitian-preserving ansatz and variational open quantum eigensolver}
\preprint{APS/123-QED}
\author{Zhong-Xia Shang}
\email{ustcszx@mail.ustc.edu.cn}
\affiliation{Hefei National Research Center for Physical Sciences at the Microscale and School of Physical Sciences,
University of Science and Technology of China, Hefei 230026, China}
\affiliation{Shanghai Research Center for Quantum Science and CAS Center for Excellence in Quantum Information and Quantum Physics,
University of Science and Technology of China, Shanghai 201315, China}
\affiliation{Hefei National Laboratory, University of Science and Technology of China, Hefei 230088, China}
\begin{abstract}
We propose a new variational quantum algorithm named Variational Open Quantum Eigensolver (VOQE) for solving steady states of open quantum systems described by either Lindblad master equations or non-Hermitian Hamiltonians. In VOQE, density matrices of mixed states are represented by pure states in doubled Hilbert space. We give a framework for building circuit ansatz which we call the Hermitian-preserving ansatz (HPA) to restrict the searching space. We also give a method to efficiently measure the operators' expectation values by post-selection measurements. We show the workflow of VOQE on solving steady states of the LMEs of the driven XXZ model and implement VOQE to solve the spectrum of the non-Hermitian Hamiltonians of the Ising spin chain in an imaginary field. 
\end{abstract}
\maketitle
\section{Introduction}
In Noisy Intermediate-Scale Quantum (NISQ) devices \cite{preskill2018quantum}, due to the lack of quantum error correction \cite{preskill2018quantum}, quantum circuits are shallow and noisy, which limits the implementations of most quantum algorithms \cite{montanaro2016quantum}. To make NISQ devices useful for practical problems, variational quantum algorithms (VQA) were proposed \cite{cerezo2021variational}. The central idea of these algorithms is evaluating quantumly and optimizing classically a cost function whose minimum (or maximum) value corresponds to the problem solution. Due to the low requirements on quantum circuits assisted by quantum error mitigation methods \cite{cai2023quantum}, VQAs have become perhaps the most promising application in the NISQ era and have attracted much attention during the past few years.

In this work, we focus on utilizing the idea of VQAs to solve problems in an important area of quantum mechanics, the open quantum systems. When a system has interactions with the environment, the behaviors of such a system can be much richer. To describe the dynamics of such systems, mixed state descriptions, non-unitary transformations, etc. need to be introduced to generalize Schrödinger's equation. Among many formaliappendixs, Lindblad Master Equation (LME) \cite{haroche2006exploring} and non-Hermitian Hamiltonian (nHH) \cite{el2018non} evolutions are rather popular and have their own successfully applicable scopes. Since the dimension of the Hilbert space can be exponentially large, solving these equations classically can be rather inefficient \cite{haroche2006exploring}, which leads to the demands on using quantum computers to solve them. There have been several proposals for open quantum systems \cite{yoshioka2020variational,endo2020variational,liu2021variational,xie2023variational,zhao2023universal}. Here, we present another new variational quantum algorithm which we call the Variational Open Quantum Eigensolver (VOQE) to solve an important topic, the steady states of open quantum systems (Hereinafter, the steady states correspond to not only those of LMEs but also the right eigenstates of this). In the following, we will first show the basic theory of VOQE which can solve the steady states of both LME and nHH, and then verify the effectiveness of VOQE on concrete problems. 

\begin{figure*}[t]
\includegraphics[width=0.8\textwidth]{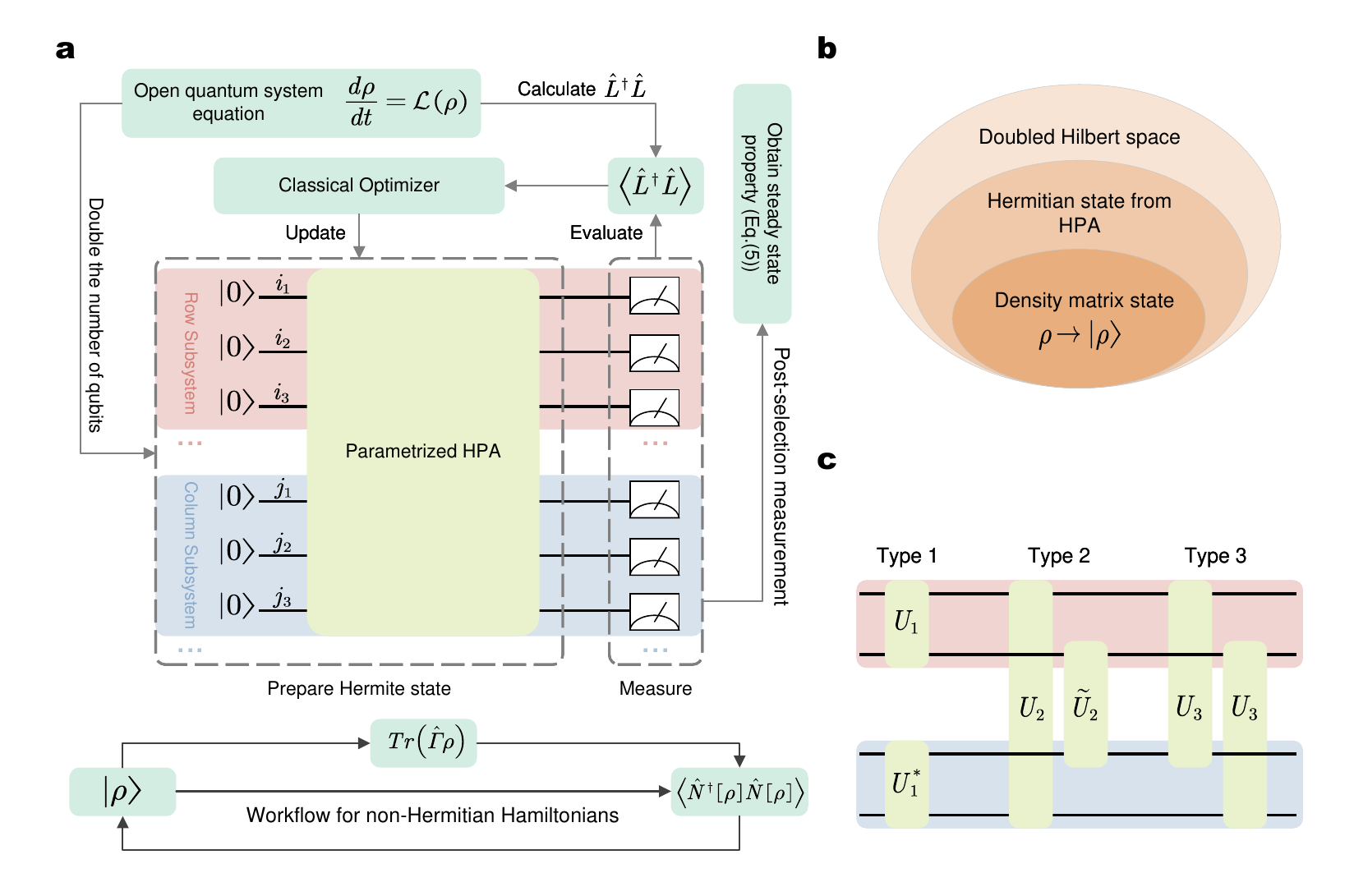}
\caption{Variational Open Quantum Eigensolver (VOQE). (a): The sketch of VOQE. VOQE uses 2n-qubit (n qubits in row subsystem and n qubits in column subsystem) parameterized HPA to solve the steady state of n-qubit open quantum system equations including LME and nHH. Equations are first transformed into the vector form to obtain the cost function operator. Next, Hermitian states from the Conjugate Ansatzes are measured to evaluate the cost function value for classical optimization. After the steady state is obtained, a post-selection method is used to obtain operators' expectation values of the state. For nHHs, unlike LME, the trace-preserving term in Eq.(\ref{nhh}) leads to non-linear equations, which makes $Tr[\Gamma\rho]$ appear in $\mathcal{N}[\rho]$. Thus, there is an additional intermediate process as shown below. (b): Relations between the whole doubled Hilbert space, Hermitian state space, and Density matrix state space. Hermitian states have the Hermiticity restriction while density matrix states not only require the Hermiticity but also the positive semi-definiteness. (c): Basic HPA blocks. There are three basic blocks when one uses single or two-qubit gates to compose an HPA. Both types share the idea of pairing to satisfy the HPA conditions Eq. \ref{lqg}. Type 1 describes the unitary transformation process while Type 2 and Type 3 simulate the non-unitary process.}\label{fig1}
\end{figure*}

LME is a rigorous quantum description of microscopic open quantum
systems assuming the Markov approximation of the environment.
An LME can be expressed as:
\begin{equation}\label{lme}
\frac{d\rho}{dt}=\mathcal{L}[\rho]=-i[H,\rho]+
\sum_i \gamma_i(F_i\rho F_i^\dag-
\frac{1}{2}\{\rho,F_i^\dag F_i\})
\end{equation}
where the Hamiltonian $H$ is the unitary part of the dynamics and $F_\mu$ are quantum
jump operators with
strength $\gamma_\mu$ describing the dissipative channels induced by the environment. For macroscopic scales, we can instead use the nHH, a semi-classical approach to encapsulate behaviors
of open quantum systems. The evolution under an nHH $H_{nh}=H-i\Gamma$
where $H$ and $\Gamma$ are Hermitian operators
can be described as:
\begin{equation}\label{nhh}
\frac{d\rho}{dt}=\mathcal{N}[\rho]=-i[H,\rho]-\{\Gamma,\rho\}+2Tr(\Gamma\rho)\rho
\end{equation}
The last term in Eq. \ref{nhh} is added to preserve the overall probability i.e. $Tr(\rho) = 1$. nHHs have rich properties such as the PT symmetry phases and the exceptional points \cite{el2018non}, which have attracted much attention in recent years. VOQE aims to solve the stead states of both Eq. \ref{lme} and Eq. \ref{nhh} i.e. $\mathcal{L}[\rho_{lss}]=0$ and $\mathcal{N}[\rho_{nss}]=0$. Note that $\mathcal{N}[\rho_{nss}]=0$ actually is the condition for eigenstates of nHHs. The basic sketch of VOQE is shown in Fig. \ref{fig1}a. In the following, we will explain details of the algorithm including the cost function, the circuit ansatz, and the way to evaluate operators' expectation values. 

\section{VOQE}
In order to give a measurable cost function for optimizations, we adopt the idea of mapping density matrices to pure states in the doubled Hilbert space $\mathcal{H}\otimes\mathcal{H}$ \cite{yoshioka2020variational,shang2024polynomial}:
\begin{equation}\label{map}
\rho=\sum_{ij}\rho_{ij}|i\rangle\langle j|\longrightarrow |\rho\rangle=
\frac{1}{C}\sum_{ij}\rho_{ij}|i,j\rangle
\end{equation}
where $C=\sqrt{\sum_{ij}|\rho_{ij}|^2}$. Note that this encoding is different from the standard purification of mixed states \cite{preskill1998lecture} used in many proposals. We call the left subsystem $\mathcal{H}$ of
$\mathcal{H}\otimes\mathcal{H}$ as the Row Subsystem(RS) and the right as the Column Subsystem(CS).
After this mapping, an operation on the density matrix $A\rho B$ is transformed into the form
$A\otimes B^T |\rho\rangle$. Following this rule, we obtain the vector representation of Eq. \ref{lme} and Eq. \ref{nhh}:
$\frac{d|\rho\rangle}{dt}=\hat{L}|\rho\rangle$ and $\frac{d|\rho\rangle}{dt}=\hat{N}[\rho]|\rho\rangle$ where $\hat{L}$ and $\hat{N}[\rho]$ are matrices (see appendix for concrete forms) acting on $|\rho\rangle$ ($\hat{N}[\rho]$ has dependence on $\rho$ which we will talk about later). The steady state $\rho_{ss}$ will satisfy the condition $C_L[|\rho_{ss}\rangle]=\langle \rho_{ss}|\hat{L^\dag}\hat{L}|\rho_{ss}\rangle=0$ for LME and $C_n[|\rho_{ss}\rangle]=\langle\rho_{ss}|\hat{N^\dag}[\rho_{ss}]\hat{N}[\rho_{ss}]|\rho_{ss} \rangle=0$ for nHH. Since the Hermitian matrices in this condition have non-negative spectra, we can thus define the cost functions as $C_L[|\rho\rangle]$ and $C_n[|\rho\rangle]$ whose minimum values 0 correspond to steady states \cite{yoshioka2020variational}.

The ansatz circuit in the doubled Hilbert space deserves a careful look. Because density matrices satisfy the Hermiticity and the positive semi-definiteness, pure states mapped from them which we will call density matrix states (DMS) only occupy part of doubled Hilbert space. An ansatz that can only be able to explore DMS has been given in dVQE \cite{yoshioka2020variational}. Here, instead, we relax the restriction of the positive semi-definiteness and give another ansatz which we will call the Hermitian-preserving ansatz (HPA) that can explore states mapped from Hermitian matrices which we will call Hermitian states satisfying $\langle i,j|\phi\rangle=\langle j,i|\phi\rangle^*$ (Fig. \ref{fig1}b). Since such ansatzes have restricted searching space and are specially designed for open quantum systems, they as problem-inspired ansatzes may have large derivations from a unitary 2-design \cite{mcclean2018barren,cerezo2021variational} and thus could have less severe barren plateau problems compared with random quantum circuits \cite{mcclean2018barren}. HPA is inspired from the similarity between the Kraus sum representation \cite{haroche2006exploring} of general quantum processes and the operator-Schmidt decomposition of unitary operators\cite{nielsen2003quantum}, which has the form:
\begin{equation}\label{lqg}
U_{HPA}=\sum_\alpha\lambda_\alpha A_{\alpha}\otimes A^*_{\alpha}
\end{equation}
where $\lambda_\alpha$ are real numbers and $A_{\alpha}$ $(A^*_{\alpha})$ are orthogonal operators bases in RS (CS), i.e. $tr[A_\alpha A_\beta^\dag]=\delta_{\alpha\beta}$. HPA Eq. \ref{lqg} is actually a representation of orthogonal matrices in real linear space spanned by Hermitian state bases (such as Hermitian states mapped from Pauli operators), thus HPA can preserve Hermitian states and is universal (see proofs in appendix). We need to mention that enlarging the searching area won't give wrong answers i.e. non-physical steady states (we give a simple proof in the appendix).

HPA can be built from 3 basic types of 2-qubit blocks (Fig. \ref{fig1}c). All the 3 blocks share the same idea of pairing gates to satisfy the condition Eq. \ref{lqg}. The first type (Type 1) has only one non-zero $\lambda_\alpha$ when written as Eq. \ref{lqg}. This type is simply the tensor product of a unitary operator in RS and its complex conjugate in CS which simulates the unitary transformations of the density matrix. Type 2 and 3 have more than one non-zero $\lambda_\alpha$ which can simulate the non-unitary dissipative transformations of the density matrix and lead to the change of density matrix eigenvalues. Here, $\tilde{U_2}$ in Type 2 is defined as $\tilde{U_2}=\sum_\alpha\lambda_\alpha^* B^*_{\alpha}\otimes A_{\alpha}^*$ acting on qubits $i_2$ and $j_1$ in order to make a pair with $U_2$ expressed as operator-Schmidt $U_2=\sum_\alpha\lambda_\alpha A_{\alpha}\otimes B_{\alpha}$ acting on qubits $i_1$ and $j_2$. Due to the pairing, both $U_1$ and $U_2$ are arbitrary. For the way of pairing in Type 3 (one $U_3$ acts on qubit $i_1$ and $j_1$ while the other acts on qubit 2 and 4), however, the form of $U_3$ has to be restricted to satisfy Eq. \ref{lqg}. As an example, the CZ gate is a typical Type 3 gate that can be $U_3$. (see details of the three types in the appendix.) 

The last segment of our algorithm uses post-selection measurements to obtain the operators' expectation values of steady states. Now suppose we have successfully found the state $|\rho_{ss}\rangle$ corresponding to the steady density matrix $\rho_{ss}$. The expectation value of an operator $O$ for $\rho_{ss}$ is $Tr[O\rho_{ss}]$ which can be expressed in terms of $|\rho_{ss}\rangle$:
\begin{equation}\label{mea}
tr[O\rho_{ss}]=\sum_i\langle i,i|O\otimes I|\rho_{ss}\rangle/\sum_i\langle i,i|\rho_{ss}\rangle
\end{equation}
To measure the right hand side of Eq.(\ref{mea}), one needs to first rotate $|\rho_{ss}\rangle$ to the eigenvector basis of $O$ and then post-select the measurement samples on all $|i,i\rangle$ bases which correspond to diagonal bases of density matrix. Suppose after measurements there are $m_i$ samples on the $|i,i\rangle$ basis, then the RHS of Eq. \ref{mea} can be estimated by:
\begin{equation}\label{sam}
\frac{\sum_i\sqrt{m_i}o_i}{\sum_i\sqrt{m_i}}
\end{equation}
where $o_i$ is $O$'s element at the $|i\rangle\langle i|$ basis. Eq. \ref{sam} is reasonable because the physical steady solution $|\rho_{ss}\rangle$ has real and positive amplitudes on $|i,i\rangle$ bases. We proved the number of required measurements to achieve an accuracy $\varepsilon $ is of order $\mathcal{O}(\eta^{-1}\varepsilon^{-4})$ where $\eta$ is the probability ratio between diagonal and non-diagonal elements of steady states. The ratio $\eta$ can vary from an exponentially small value ($2^{-n}$ when the density matrix corresponds to $|+\rangle^{\otimes n}$) to 1 for the maximally mixed state. Roughly speaking, a small purity will lead to a higher $\eta$ and thus a smaller sampling cost. Thus, this ratio is acceptable for most problem models due to their dissipative nature (See details about this method and its measurement cost in the appendix). Note that one can also use Hadamard tests \cite{datta2008quantum} and swap tests \cite{barenco1997stabilization} for evaluating Eq. \ref{mea}, which however, might be unfriendly for NISQ devices.
\begin{figure*}[tbp]
\centering
\includegraphics[width=0.99\textwidth]{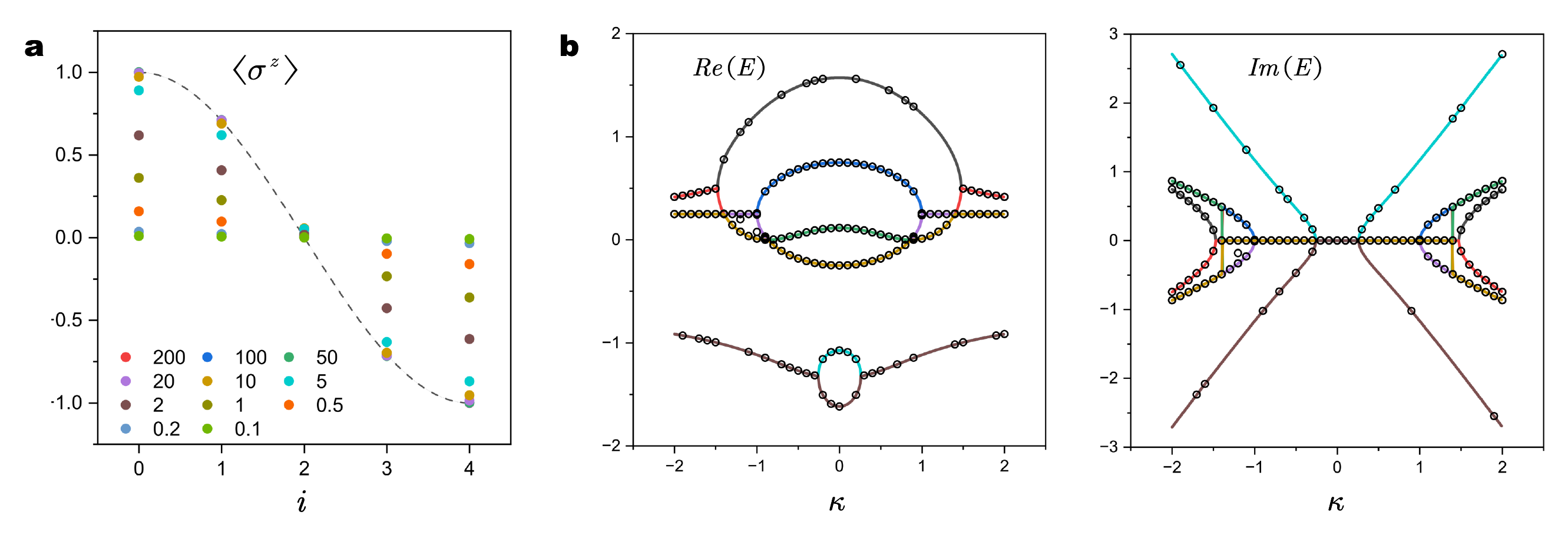}
\caption{Numerical experiments of VOQE. (a): The steady states of the LME of the driven open XXZ model. We set $\Delta=1$ and turn $\epsilon$ from 200 to 0.1. There will appear cosine spin profile $\langle\sigma_i^z\rangle=cos(\pi\frac{i-1}{n-1})$ as the $\epsilon$ increases to a large value. The problem size is 5-qubit and a 10-qubit HPA is used for training. (b): The complex spectrum of the nHH of the Ising spin chain in an imaginary field. The real part and the imaginary part are plotted respectively. The solid lines form the exact complex spectrum and the points are obtained from VOQE. We set $\lambda=0.5$ and turn $\kappa$ from -2 to 2, the spectrums of the Hamiltonians of the model are complex except for the PT-symmetry phases. For each $(\lambda,\kappa)$ setting, we run VOQE 30 times to make sure the majority of the spectrum is covered. The problem size is 3-qubit.}\label{fig2}
\end{figure*}

These three segments compose the whole structure of VOQE as shown in Fig. \ref{fig1}a. In general, for an $n$-qubit open quantum equations, we can build a parameterized 2$n$-qubit HPA to train the steady states and use the measurement protocol to obtain steady state information. One thing to mention here is that for nHHs, unlike LME, the trace-preserving term in Eq.(\ref{nhh}) leads to non-linear equations, which makes $Tr[\Gamma\rho]$ appear in $\mathcal{N}[\rho]$. Thus, there is an additional intermediate process for evaluating $Tr(\Gamma\rho)$. Also, since only Type 1 circuits are needed because nHHs won't lead to mixed states, an $n$-qubit system that prepares trial states $|\psi\rangle$ and $|\psi^*\rangle$ at different times is enough for getting the cost functions.
\begin{figure}[hbp]
\centering
\includegraphics[width=0.48\textwidth]{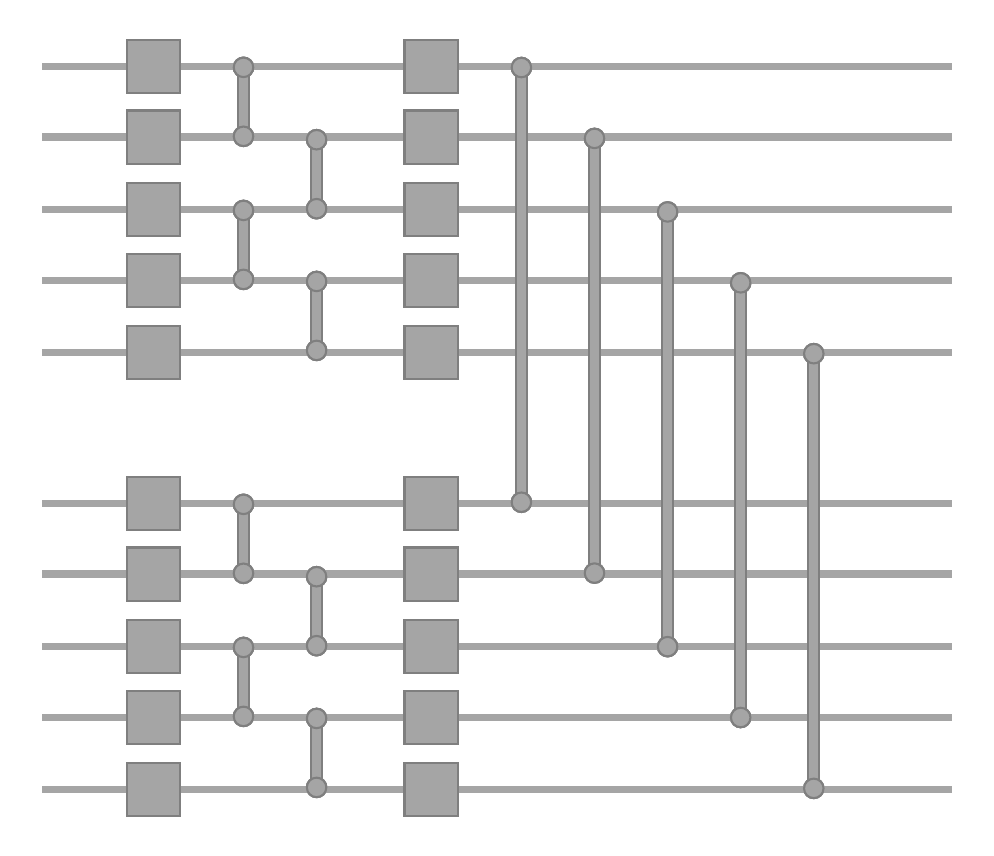}
\caption{A layer of the parameterized HPA for the numerical experiments on the driven open XXZ model. During the experiments, we fixed the layer depth to be 1 with additional single-qubit parameterized gates appended at the end of the ansatz. Here, depth means the number of HPA layers shown in this figure.}\label{fig3}
\end{figure}
\section{Numerical experiments}

To verify the effectiveness of VOQE, we run numerical experiments for specific problems. One is the LME of the driven open XXZ model \cite{prosen2011exact} with the Hamiltonian $H=\sum_i \Delta \sigma_i^z\sigma_{i+1}^z+2 \sigma_i^+\sigma_{i+1}^-+2\sigma_i^-\sigma_{i+1}^+$ of open boundarys and two jump channels $F_1=\sigma_1^+$ and $F_2=\sigma_n^-$ of strength $\epsilon$. The parameterized circuit for training is composed of Type 1 and Type 3 gates as shown in Fig. \ref{fig3}. Type 1 gates contain layered single-qubit parameterized gates with the form $e^{-i\theta_x X}e^{-i\theta_y Y}e^{-i\theta_z Z}$ and fixed CZ gates. Note that a single-qubit gate with parameters $\{\theta_x,\theta_y,\theta_z\}$ in RS is paired with a single-qubit gate in the CS with parameters $\{-\theta_x,\theta_y,-\theta_z\}$ to satisfy the HEA condition. Type 3 gates are CZ gates. For the steady states of this model in the isotropic case $\Delta=1$, there will appear cosine spin profile $\langle\sigma_i^z\rangle=cos(\pi\frac{i-1}{n-1})$ as the $\epsilon$ increases to a large value. By turning $\epsilon$ from 200 to 0,1, we observe such behaviors in our experiment by variationally preparing the steady states and using Eq. \ref{mea} to obtain spins' expectation values of interest. The results can be found in Fig. \ref{fig2}a. (The convergence of the cost functions with respect to iteration steps can be found in the appendix.) The other problem is the nHH of the Ising spin chain in an imaginary field $H=-\frac{1}{2}\sum_i(\sigma_i^z+\lambda \sigma_i^x\sigma_{i+1}^x+i\kappa\sigma_i^x)$ with periodic boundary \cite{castro2009spin}. The parameterized circuit is composed of only 2-qubit gates from Type 1 (with a depth of three ignoring Type 3 CZ gates in Fig. \ref{fig3}) since nHHs won't lead to mixed states. Since all eigenstates satisfy $C_n[|\rho\rangle]=0$, we can use the algorithm to draw the spectrum of nHHs by repeated experiments. We set $\lambda=0.5$ and turn $\kappa$ from -2 to 2, the spectrums of the Hamiltonians of the model are complex except for the PT-symmetry phases. We recover the spectrums in Fig. \ref{fig2}b. Note that if one wants to find specific eigenstates, penalty terms and pre-optimizations \cite{xie2023variational} can be added. The classical optimization method used throughout the experiments is the BFGS algorithm assisted by the idea of adiabatic variational optimizing \cite{garcia2018addressing,harwood2022improving} where the approximated ground state optimized at a point is chosen as the initial state of points close to it.

\section{Comparison with related works}
Now, we want to give a discussion on the comparison between VOQE and other variational quantum algorithms for open quantum systems in Ref. \cite{yoshioka2020variational,endo2020variational,liu2021variational,xie2023variational,zhao2023universal}. First of all, to the best of our knowledge, compared with these mentioned works, VQOE is the first variational quantum algorithm that can solve steady-state problems of both LME and nHHs in a unified framework. The algorithm proposed in Ref. \cite{endo2020variational} focuses on the variational simulations of dynamics of open quantum systems rather than the steady state problems. Also, the way they encode density matrices is by purification \cite{preskill1998lecture} rather than vectorization used in VOQE. In Ref. \cite{liu2021variational}, while the authors propose a VQA for the steady states of LME, the purification encoding makes them have to introduce swap tests \cite{buhrman2001quantum} to evaluate a non-linear cost function. In contrast, by using the vectorization encoding, Ref. \cite{yoshioka2020variational} and our work are able to use the expectation values of $L^\dag L$ as natural cost functions that can be easily evaluated by the direct operator averaging method \cite{mcclean2016theory}. Compared with Ref. \cite{yoshioka2020variational}, HEA proposed in VOQE can have a more flexible structure and thus has a potentially better friendliness for NISQ hardware. Also, we gave an alternative measurement strategy for obtaining steady-state information. Ref. \cite{xie2023variational,zhao2023universal} focus on using the variational framework to solve nHH problems. In Ref. \cite{xie2023variational}, the authors use the variances of nHH energy as the cost functions and can not be directly generalized to LME. In Ref. \cite{zhao2023universal}, the authors gave a variational quantum algorithm for the eigenvalues of nHHs based on diagonalizations, which requires complicated quantum circuits that are unfriendly for NISQ devices.
\section{Summary and outlook}
In summary, we have presented a variational quantum algorithm for solving the steady states of LMEs and nHHs. density matrices are mapped to pure states in the doubled Hilbert space for measurable cost functions. We constructed the Hermitian-preserving ansatz to restrict the searching space. We want to mention that the applications of such Hermitian-preserving ansatzes should not be restricted to VOQE and can be further investigated. We also gave a post-selection measurement method to evaluate operators' expectation values of the steady states. Our algorithms are tested for specific problems and the results coincide with the theoretical predictions. We hope this work will show a future application for NISQ devices and motivate people to utilize the idea of variational quantum algorithms for solving various problems.

We used the Qulacs \cite{suzuki2021qulacs} for our numerical experiments.

\begin{acknowledgements}
This work is supported by the National Natural Science Foundation of China (No. 91836303 and No. 11805197), the National Key R$\&$D Program of China, the Chinese Academy of Sciences, the Anhui Initiative in Quantum Information Technologies, and the Science and Technology Commission of Shanghai Municipality (2019SHZDZX01). The authors would like to thank MC Chen and CY Lu for their insightful advice.
\end{acknowledgements}
\nocite{*}
\bibliography{ref.bib}
\onecolumngrid
\appendix
\section{Concrete forms of $\hat{L}$ and $\hat{N}[\rho]$}

\begin{equation}
\begin{aligned}
&\hat{L}=(-i(H\otimes I-I\otimes H^T)+\sum_i D[\Gamma_i])\\
&\text{where} \quad D[\Gamma_i]=F_i\otimes F_i^*-\frac{1}{2}F_i^\dag F_i\otimes I-I\otimes\frac{1}{2}F_i^TF_i^*
\end{aligned}
\end{equation}

\begin{equation}
\hat{N}[\rho]=(-i(H\otimes I-I\otimes H^T)-(\Gamma\otimes I+I\otimes \Gamma^T)+2Tr[\Gamma\rho]I\otimes I
\end{equation}

\section{Uniqueness of $|\rho_{ss}\rangle$}

We assume the condition is there is only one unique steady density matrix of a LME. However, the question is if the uniqueness will still hold if we enlarge the density matrix states to the Hermitian states since there may exist other non-density matrix states that are eigenvectors of the Liouvillian operator $\hat{L}$ of the LME with zero eigenvalues.

Suppose there is not only one unique steady density matrix state $|\rho_{ss}\rangle$ but also one Hermitian steady state $|\rho_h\rangle$.
We can decompose $|\rho_h\rangle$ into:
\begin{equation}
|\rho_h\rangle=c_1|\rho_1\rangle + c_2|\rho_2\rangle
\end{equation}
where $c_1$ and $c_2$ are real numbers and $|\rho_1\rangle$ and $|\rho_2\rangle$ are density matrix state. $|\rho_1\rangle$ and $|\rho_2\rangle$ can further be decomposed as:
\begin{equation}
|\rho_1\rangle= |\rho_{ss}\rangle+ |\rho_1'\rangle,\quad |\rho_2\rangle= |\rho_{ss}\rangle+ |\rho_2'\rangle
\end{equation}
Thus, we have:
\begin{equation}
|\rho_h\rangle=(c1+c2)|\rho_{ss}\rangle+c_1|\rho_1'\rangle+c_2|\rho_2'\rangle
\end{equation}
Due to the unique steady density matrix state condition, $|\rho_1'\rangle$ and $|\rho_2'\rangle$ must be linear combinations of eigenvectors of the Liouvillian operator $\hat{L}$ of a LME with nonzero eigenvalues.
Therefore, $|\rho_h\rangle$ can't be a steady Hermitian state which proves VOQE won't give a wrong answer. nHH won't have this issue since only Type 1 circuits are required.

\section{HPA}

A completely positive transformation(CPT) can be written as the Kraus sum
\begin{equation}\label{c1}
\rho\rightarrow\sum_\alpha M_\alpha \rho M_\alpha^\dag
\end{equation}
If we only want to keep Hermiticity of the matrix, Eq.(\ref{c1}) can be adjusted to 
\begin{equation}\label{c2}
\rho\rightarrow\sum_\alpha \eta_\alpha M_\alpha \rho M_\alpha^\dag
\end{equation}
where $\eta_\mu$ is real. To keep the trace of the matrix one, the following equation must be obeyed
\begin{equation}\label{c3}
\sum_\alpha\eta_\alpha M_\alpha^\dag M_\alpha=I
\end{equation}
However, in order to keep a HPA described as 
\begin{equation}\label{lq}
U_{HPA}=\sum_\alpha\lambda_\alpha A_{\alpha}\otimes A^*_{\alpha}
\end{equation}
to be unitary, it must obey
\begin{equation} \label{c4}
\sum_{\alpha\beta} \lambda_\alpha\lambda_\beta A_\alpha^\dag A_\beta\otimes A_\alpha^T A_\beta^*=I
\end{equation}
Eq.(\ref{c3}) and Eq.(\ref{c4}) are the same condition if and only if the HPA is composed of only Type 1 circuit blocks. For other types, HPA and Kraus sum are not one-to-one correspondence.

A universal HPA form Eq.(\ref{lq}) can be obtained by considering orthogonal matrices of linear space spanned by Hermitian state bases. An orthogonal matrix $U_{HPA}$ in this space can be expressed as diagonal form
\begin{equation} \label{c5}
U_{HPA}=\sum_\beta \zeta_\beta |\Psi^\beta\rangle\langle\Psi^\beta|
\end{equation}
where $\zeta_\beta=\pm1$ and $|\Psi^\beta\rangle=\sum_{ij}\Psi^\beta_{ij}|i,j\rangle$ satisfy the Hermitian state condition $\Psi^\beta_{ij}={\Psi^\beta_{ji}}^*$. The elements of the HPA satisfy
\begin{equation} \label{c7}
\begin{split}
&M_{ik,jl=}\langle i,j|U_{HPA}|k,l\rangle=\sum_\beta \zeta_\beta\Psi^\beta_{ij}{\Psi^\beta_{kl}}^*\\&
=\sum_\beta \zeta_\beta(\Psi^\beta_{ji}{\Psi^\beta_{lk}}^*)^*=\langle j,i|U_{HPA}|l,k\rangle^*=M_{jl,ik}^*
\end{split}
\end{equation}
Eq.(\ref{c7}) is the necessary and sufficient condition of a unitary operator to be a HPA. We see $M$ is a Hermitian matrix (by treating $ik$ as row index and $jl$ as column index). By diagonalizing $M$, we have $M=T\lambda T^\dag$, where $T$ is unitary and $\lambda$ is diagonal with real diagonal entries $\lambda_\alpha$. Now we can express Eq.(\ref{c5}) as 
\begin{eqnarray}
&&U_{HPA}=\sum_{ijkl}M_{ik,jl}|i,j\rangle\langle k,l|
\nonumber\\&&=\sum_{ijkl\alpha}T_{ik\alpha}\lambda_{\alpha}T^\dag_{\alpha jl}|i\rangle\langle k|\otimes |j\rangle\langle l|
\nonumber\\&&=\sum_{\alpha}\lambda_{\alpha}(\sum_{ik} T_{ik\alpha} |i\rangle\langle k|)\otimes(\sum_{jl} T^*_{jl\alpha}|j\rangle\langle l|)
\nonumber\\&&=\sum_{\alpha}\lambda_{\alpha}A_\alpha\otimes A^*_\alpha
\end{eqnarray}
It is easy to check that $A_\alpha=\sum_{ik} T_{ik\alpha}$ are orthonormal operator bases, thus we have proved Eq.(\ref{lq}). The proof process is similar to the operator-Schmidt decomposition\cite{nielsen2003quantum} where single value decomposition(SVD) replaces the diagonalization process.

\section{HPA types}

The first type of Eq.(\ref{lq}) corresponds to only one non-zero $\lambda_\alpha$. This type is simply the tensor product of a unitary operator $U_1$ in RS and its complex conjugate in CS which simulates the unitary transformation of density matrix.
\begin{equation}
U_{T1}=U_1\otimes U_1^*
\end{equation}

For the second type, it is easy to check:
\begin{equation}
U_{T2}=U_2\otimes\tilde{U_2}=\sum_{\alpha\beta}\lambda_\alpha\lambda_\beta (A_{\alpha}\otimes B^*_{\beta})\otimes (A_{\beta}^*\otimes B_{\alpha})
\end{equation}
one can further prove:
\begin{eqnarray}
&&\langle i,j|U_{T2}|k,l\rangle=\sum_{\alpha\beta}\lambda_\alpha\lambda_\beta(A_{\alpha}\otimes B^*_{\beta})_{ik}\otimes (A_{\beta}^*\otimes B_{\alpha})_{jl}
\nonumber\\&&=\sum_{\alpha\beta}\lambda_\alpha\lambda_\beta(A_{\alpha}\otimes B^*_{\beta})_{ik}(A_{\beta}^*\otimes B_{\alpha})_{jl}
\nonumber\\&&=\sum_{\alpha\beta}\lambda_\alpha\lambda_\beta(A_{\beta}\otimes B_{\alpha}^*)^*_{jl}(A^*_{\alpha}\otimes B_{\beta})^*_{ik}
\nonumber\\&&=\sum_{\beta\alpha}\lambda_\beta\lambda_\alpha(A_{\beta}\otimes B_{\alpha}^*)^*_{jl}(A^*_{\alpha}\otimes B_{\beta})^*_{ik}=\langle j,i|U_{T2}|l,k\rangle^*
\end{eqnarray}
which exactly satisfies the condition Eq.(\ref{c7}). Thus $U_{T2}$ is a HPA block.

The $U_3$ in type 3 has restricted form which can be directly obtained from Eq.(\ref{c7}):
\begin{eqnarray}
\begin{bmatrix}
M_{0000}&M_{0001}&M_{0001}^*&M_{0101}\\
M_{0010}&M_{0011}&M_{0110}&M_{0111}\\
M_{0010}^*&M_{0110}^*&M_{0011}^*&M_{0111}^*\\
M_{1010}&M_{1011}&M_{1011}^*&M_{1111}
\end{bmatrix} 
\end{eqnarray}
from which one can obtain the Type 3:
\begin{equation}
U_{T3}=U_3\otimes U_3
\end{equation}
Thus, $U_3$ doesn't need a pairing procedure since $U_3$ itself has satisfied the condition.

\section{Post-selection measurement}

In this appendix we show how to evaluate Eq.(\ref{mea}) by measurements and give the measurement cost of it. We first assume that $O$ is diagonal in $|i\rangle$ basis, i.e. $O_{ij}=o_i\delta_{ij}$. Then Eq.(\ref{mea}) can be rewritten as
\begin{eqnarray}\label{e1}
&&\sum_i\langle i,i|O\otimes I|\rho_{ss}\rangle/\sum_i\langle i,i|\rho_{ss}\rangle
\nonumber\\&&=\sum_{ikl}\rho'_{sskl}\langle i|O|k\rangle\delta_{il}/\sum _i\rho'_{ssii}
\nonumber\\&&=\sum_{ikl}\rho'_{sskl}o_i\delta_{ik}\delta_{il}/\sum _i\rho'_{ssii}
\nonumber\\&&=\sum_{i}\rho'_{ssii}o_i/\sum _i\rho'_{ssii}
\end{eqnarray}
$\rho'_{ssii}=\rho_{ssii}/C$ are real and nonnegative because the steady state corresponds to a physical density matrix, which means it can be evaluated by measurements. Consider that we repeat the measurements for totally $M$ times. If $m_i$ samples are obtained on the $|i,i\rangle$ basis and $\sum_i m_i=m$, then the post-selection efficiency is $\eta=\frac{m}{M}$ which depends on the probability ratio between diagonal and non-diagonal elements of steady states. For many dissipation models, non-diagonal elements decay to near zero, thus $\eta$ are acceptable. Eq.(\ref{e1}) can be evaluated by post-selection and post-processing:
\begin{equation}\label{e2}
\sum_{i}\rho'_{ssii}o_i\approx \frac{\sum_i\sqrt{\frac{m_i}{m}}o_i}{\sum_i\sqrt{\frac{m_i}{m}}}=\frac{\sum_i\sqrt{m_i}o_i}{\sum_i\sqrt{m_i}}
\end{equation}
The variance of the right hand side of Eq.(\ref{e2}) is
\begin{equation}
Var[\frac{\sum_i\sqrt{m_i}o_i}{\sum_i\sqrt{m_i}}]=\frac{Var[O]}{\sum_i\sqrt{m_i}}\leq \frac{Var[O]}{\sqrt{m}}=\frac{Var[O]}{\sqrt{\eta M}}
\end{equation}
Thus the measurement cost we need to achieve a variance of $\epsilon^2$ in the worst case is
\begin{equation}
M(\epsilon)\approx \frac{1}{\eta}(\frac{Var[O]}{\epsilon^2})^2
\end{equation}
For general $O$, we need to decompose them on different measurement bases (Pauli bases) $O=\sum_{\gamma=1}^K O_{\gamma}$ to evaluate the expectation value of each part individually as discussed in Ref.\cite{mcclean2016theory} and the similar result can be obtained
\begin{equation}
M(\epsilon)\approx K(\frac{\sum_\gamma Var[O_\gamma]/\sqrt{\eta_\gamma}}{\epsilon^2})^2
\end{equation}
where $\eta_\gamma$ is the efficiency of the steady state in the diagonal basis of $O_\gamma$.

\section{Convergence of cost functions in numerical experiments}

\begin{figure*}[htbp]
\centering
\includegraphics[width=0.7\textwidth]{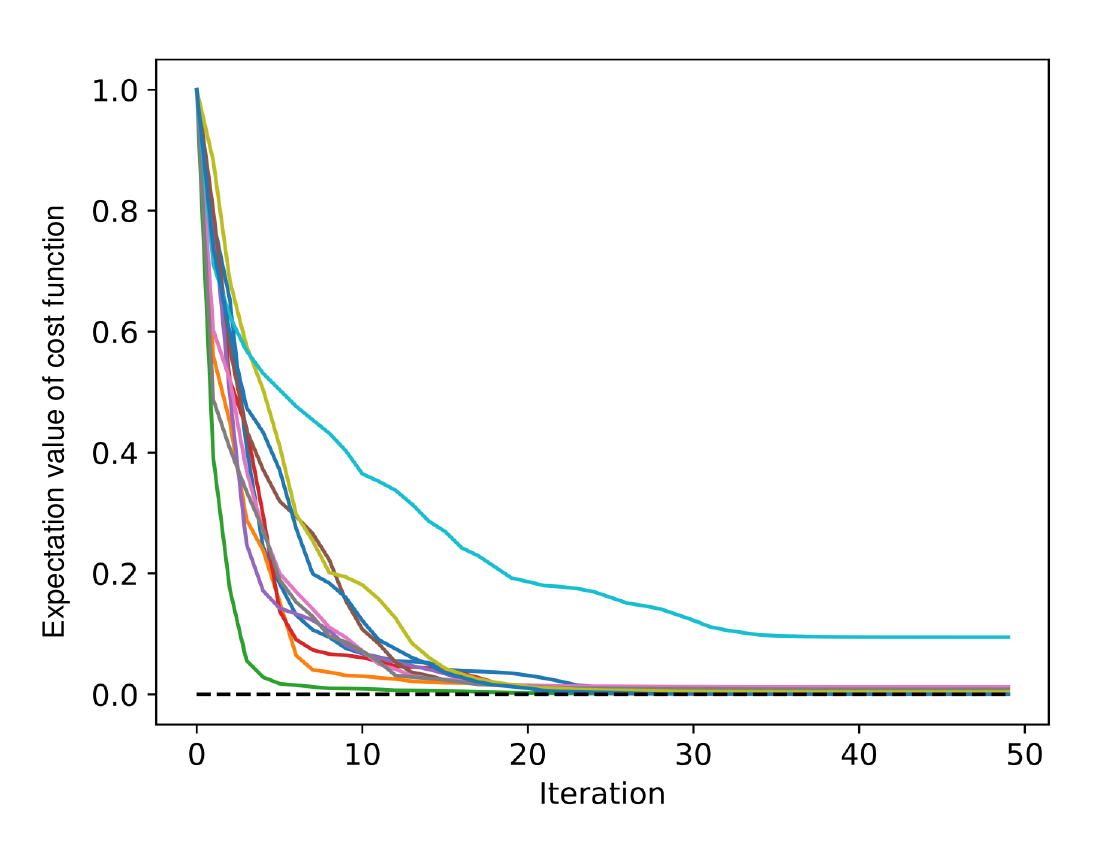}
\caption{Convergence of the cost functions of the driven open XXZ model with $\epsilon=1$ under random initialization as functions of iteration steps. For each cost function, we have re-scaled its range within $[0,1]$ to have a better presentation. The starting point (initial parameters) of each curve is chosen randomly. The optimizer is chosen to be the BFGS optimizer. }\label{fig4}
\end{figure*}
\end{document}